\documentclass{aa}  
\usepackage{graphicx}
\usepackage{txfonts}
\usepackage{xcolor}
\usepackage{hyperref}
\hypersetup{colorlinks, linkcolor=blue, citecolor=blue, urlcolor=blue}
\usepackage{amsmath}
\usepackage{ulem}
\usepackage{soul}
\usepackage{enumitem}% http://ctan.org/pkg/enumitem
\newcommand{\te}{t_{\rm E}}
\newcommand{\thetae}{\theta_{\rm E}}

\newcommand{\pie}{\pi_{\rm E}}

\newcommand{\dl}{D_{\rm L}}

\definecolor{brown}{rgb}{0.59, 0.29, 0.0}
\definecolor{darkgreen}{rgb}{0.0, 0.42, 0.24}
\definecolor{darkblue}{rgb}{0.01, 0.31, 0.59}
\definecolor{darkblue}{rgb}{0.0, 0.25, 0.42}
\definecolor{blue}{rgb}{0.0,0.0,1.0}
\definecolor{green}{rgb}{0.0,1.0,0.0}

\begin{document} 

\title{KMT-2021-BLG-1898: Planetary microlensing event involved with binary source stars}

\author{
% leading author -----------------------------
     Cheongho Han\inst{\ref{01}} 
\and Andrew Gould\inst{\ref{02},\ref{03}} 
\and Doeon~Kim\inst{\ref{01}} 
\and Youn~Kil~Jung\inst{\ref{04}} 
% KMTNet ---------------------------
\and Michael~D.~Albrow\inst{\ref{05}} 
\and Sun-Ju~Chung\inst{\ref{04}} 
\and Kyu-Ha~Hwang\inst{\ref{04}} 
\and Chung-Uk~Lee\inst{\ref{04}} 
\and Yoon-Hyun~Ryu\inst{\ref{04}} 
\and In-Gu~Shin\inst{\ref{04}} 
\and Yossi~Shvartzvald\inst{\ref{07}} 
\and Jennifer~C.~Yee\inst{\ref{08}} 
\and Weicheng Zang\inst{\ref{06}} 
\and Sang-Mok~Cha\inst{\ref{04},\ref{09}} 
\and Dong-Jin~Kim\inst{\ref{04}} 
\and Seung-Lee~Kim\inst{\ref{04}} 
\and Dong-Joo~Lee\inst{\ref{04}}
\and Yongseok~Lee\inst{\ref{04}} 
\and Byeong-Gon~Park\inst{\ref{04}} 
\and Richard~W.~Pogge\inst{\ref{03}}
\\
(The KMTNet Collaboration),\\
}

\institute{
       Department of Physics, Chungbuk National University, Cheongju 28644, Republic of Korea  \\ \email{cheongho@astroph.chungbuk.ac.kr}              \label{01} 
\and   Max Planck Institute for Astronomy, K\"onigstuhl 17, D-69117 Heidelberg, Germany                                                                \label{02} 
\and   Department of Astronomy, The Ohio State University, 140 W.  18th Ave., Columbus, OH 43210, USA                                                  \label{03} 
\and   Korea Astronomy and Space Science Institute, Daejon 34055, Republic of Korea                                                                    \label{04} 
\and   University of Canterbury, Department of Physics and Astronomy, Private Bag 4800, Christchurch 8020, New Zealand                                 \label{05} 
\and   Department of Astronomy, Tsinghua University, Beijing 100084, China                                                                             \label{06} 
\and   Department of Particle Physics and Astrophysics, Weizmann Institute of Science, Rehovot 76100, Israel                                           \label{07} 
\and   Center for Astrophysics~|~Harvard \& Smithsonian, 60 Garden St., Cambridge, MA 02138, USA                                                       \label{08}
\and   School of Space Research, Kyung Hee University, Yongin, Kyeonggi 17104, Republic of Korea                                                       \label{09}  
%\\
%\email{cheongho@astroph.chungbuk.ac.kr}   
}
\date{Received ; accepted}

% \abstract{}{}{}{}{} 
% 5 {} token are mandatory
\abstract
{}
{
The light curve of the microlensing event KMT-2021-BLG-1898 exhibits a short-term central anomaly
with double-bump features that cannot be explained by the usual binary-lens or binary-source 
interpretations.  With the aim of interpreting the anomaly, we analyze the lensing light curve 
under various sophisticated models.
}
{
We find that the anomaly is explained by a model, in which both the lens and source are binaries
(2L2S model).  For this interpretation, the lens is a planetary system with a planet/host mass 
ratio of $q\sim 1.5\times 10^{-3}$,  and the source is a binary composed of a turn off or a 
subgiant star and a mid K dwarf.  The double-bump feature of the anomaly can also be depicted by 
a triple-lens model (3L1S model), in which the lens is a planetary system containing two planets.  
Among the two interpretations, the 2L2S model is favored over the 3L1S model not only because it 
yields a better fit to the data, by $\Delta\chi^2=[14.3$--18.5], but also the Einstein radii 
derived independently from the two stars of the binary source result in consistent values.  
According to the 2L2S interpretation, KMT-2021-BLG-1898 is the third planetary lensing event 
occurring on a binary stellar system, following MOA-2010-BLG-117 and KMT-2018-BLG-1743.
}
{
Under the 2L2S interpretation, we identify two solutions resulting from the close-wide degeneracy 
in determining the planet-host separation.  From a Bayesian analysis, we estimate that the planet 
has a mass of $\sim 0.7$--0.8~$M_{\rm J}$, and it orbits an early M dwarf host with a mass of 
$\sim 0.5~M_\odot$. The projected planet-host separation is $\sim 1.9$~AU and $\sim 3.0$~AU 
according to the close and wide solutions, respectively.
} 
% conclusions heading (optional), leave it empty if necessary {}
{}


\maketitle

\section{Introduction}\label{sec:one}

Microlensing planets are detected and characterized by analyzing short-term anomalies in lensing
light curves induced by planets. In most cases, planetary anomalies are well described by a 
binary-lens single-source (2L1S) model, in which the lens is composed of two masses (planet and 
its host) and the source is a single star. However, some planetary signals are deformed from the 
2L1S form due to various causes. The first major cause for such a deformation is the existence 
of an extra lens component.  This additional lens component can be a second planet, as in the 
cases of OGLE-2006-BLG-109 \citep{Gaudi2008, Bennett2010}, OGLE-2012-BLG-0026 \citep{Han2013}, 
and OGLE-2018-BLG-1011 \citep{Han2019}, or a binary companion to the host, as in the cases of 
OGLE-2007-BLG-349 \citep{Bennett2016}, KMT-2020-BLG-0414 \citep{Zang2021}, OGLE-2016-BLG-0613 
\citep{Han2017}, and OGLE-2018-BLG-1700 \citep{Han2020}.  Another major cause of a deformation 
is the binarity of the source, as illustrated by the lensing events MOA-2010-BLG-117 
\citep{Bennett2018} and KMT-2018-BLG-1743 \citep{Han2021a}. For the lensing event KMT-2019-BLG-1715, 
an even more complicated model with three lens masses (binary stars and a planet) and two source 
stars is needed to explain the deformed anomalies in the lensing light curve \citep{Han2021c}.

Interpreting a deformed planetary anomaly is a difficult task because of the complexity of
modeling with the increased number of lensing parameters. Even in the simplest case of a 
single planetary event, seven basic parameters are needed to describe the observed light curve. 
Considering an extra lens component (3L1S) or a source component (2L2S) in modeling requires 
one to include multiple extra parameters in addition to the basic parameters, and this results 
in the complexity in modeling.  We give an explanation of the lensing parameters for the 
individual models (2L1S, 3L1S, and 2L2S) in the analysis part of the paper. As a result, some 
planetary signals with complex structures may be missed due to the difficulty of modeling. 
For example, the planetary nature of the lensing events OGLE-2018-BLG-1700, KMT-2018-BLG-1743, 
and KMT-2019-BLG-1715 had not been known before they were reanalyzed with complex models 
from the systematic reinvestigation of anomalous events with no presented lensing models.  
Considering that planet statistics are based on the detection efficiency, which is estimated 
as the ratio of lensing events with detected planetary signals to the total number of lensing 
events, missing planets would lead to erroneous planet statistics such as the planet frequency 
and demographic distribution.

In this work, we report a planet found from the analysis of the microlensing event
KMT-2021-BLG-1898. A short-term anomaly appeared near the peak of the lensing light curve, 
but it cannot be described by the usual binary-lens or binary-source (1L2S) model.  We 
investigate various causes for the deformation of the anomaly to reveal the nature of the 
anomaly.

We present the analysis of the planetary event according to the following organization. In 
Sect.~\ref{sec:two}, we give an explanation of the data, including observations, facilities, 
and data reductions. In Sect.~\ref{sec:three}, we describe the detailed features of the anomaly, 
and demonstrate the difficulty of describing the anomaly with the usual 2L1S or 1L2S models.  
We check the feasibility of describing the anomaly with various sophisticated models and present 
the analyses. We estimate the angular Einstein radius by specifying the source in 
Sect.~\ref{sec:four}, and estimate the physical parameters of the planet system in 
Sect.~\ref{sec:five}. We summarize the results and conclude in Sect.~\ref{sec:six}.

\section{Observations and data}\label{sec:two}

The source of the lensing event KMT-2021-BLG-1898 lies toward the Galactic bulge field with 
equatorial coordinates (R.A., decl.)$_{\rm J2000}=$(17:42:46.05, -27:22:33.02), which correspond 
to the galactic coordinates $(l, b) = (1^\circ\hskip-2pt .000, 1^\circ\hskip-2pt .353)$. Due to 
the proximity of the source to the Galactic center, the extinction toward the field, $A_I=3.56$, 
is fairly high.  The magnitude of the baseline object is $i=20.64$ as reported in DECam catalog of 
\citet{Schlafly2018}.

The lensing event was detected from the survey conducted by the Korea Microlensing Telescope 
Network \citep[KMTNet;][]{Kim2016} team during the 2021 bulge season.  The event reached a 
relatively high magnification, $A_{\rm peak}\sim 63$, at the peak on 2021 July 26 
(${\rm HJD}^\prime\equiv {\rm HJD}-2450000\sim 9421.6$).  The peak region of a high-magnification 
is susceptible to deviations induced by a planetary companion \citep{Griest1998}, and a short-term 
anomaly was actually found in the data collected by the KMTNet survey.  Had the event been alerted 
before the peak, the anomaly could have been densely resolved by follow-up observations utilizing 
multiple telescopes, but no follow-up observation could be conducted because the event was alerted 
just after the peak.  Furthermore, there are no data from the other surveys because the telescope 
of the OGLE survey was shut down in the 2021 season and the MOA survey did not find the event.  We 
give a detailed description of the anomaly features in the following section.

% Figure 1 ------------------------------------------------------
\begin{figure}[t]
\includegraphics[width=\columnwidth]{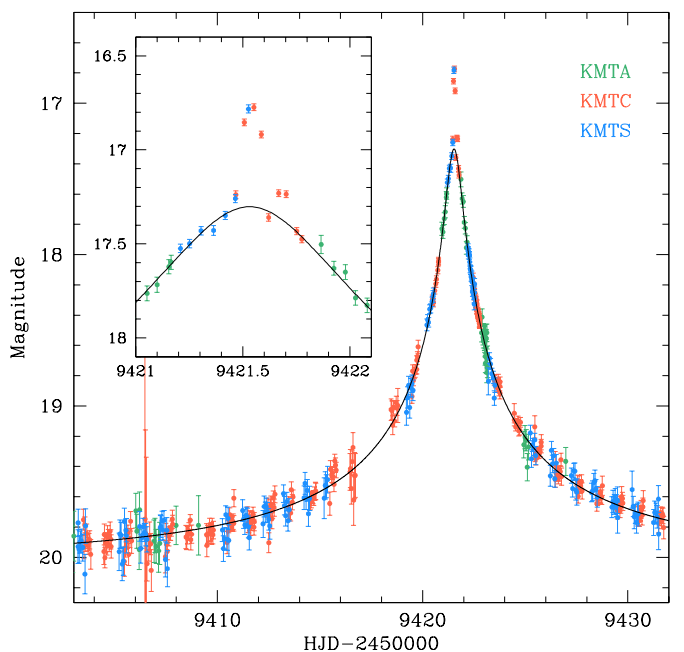}
\caption{
Light curve of KMT-2021-BLG-1898 constructed by combining the three data sets of the KMTNet 
telescopes: KMTA, KMTC, and KMTS. The inset shows the zoom-in view of the peak region exhibiting 
an anomaly feature. The curve drawn over the data point is a 1L1S model obtained by excluding the 
data points in the anomaly region.  The colors of the data points are set to mach those of the 
telescopes marked in the legend.
}
\label{fig:one}
\end{figure}
% --------------------------------------------------------------

Observations of the event were done utilizing the three telescopes of the KMTNet survey. The
telescopes are identical with a 1.6~m aperture, and they are globally distributed in the three
continents of the Southern Hemisphere for continuous coverage of lensing events. The sites of the
individual telescopes are the Siding Spring Observatory (KMTA) in Australia, the Cerro Tololo
Inter-American Observatory (KMTC) in South America, and the South African Astronomical Observatory 
(KMTS) in Africa. Each telescope is equipped with a wide-field camera yielding 4~deg$^2$ field of 
view. The lensing event is in the KMT18 field, toward which observations were conducted with a 
1~hr cadence. Images of the field were mostly acquired in the $I$ band, but 9\% of the images 
were obtained in the $V$ band for the source color measurement.  We discuss the procedure of the 
source color measurement in Sect.~\ref{sec:four}.

Photometry of the lensing event was done using the KMTNet pipeline \citep{Albrow2009}, which 
applies the difference image method \citep{Tomaney1996, Alard1998}. For a subset of the KMTC 
$I$- and $V$-band data, additional photometry was done using the pyDIA software \citep{Albrow2017} 
to construct a color-magnitude (CMD) of stars around the source and estimate the source location 
on the CMD.  For the data used in the analysis, we readjust the error bars following the routine 
described in \citet{Yee2012}, that is, $\sigma=k(\sigma_{\rm min}+\sigma_0)^{1/2}$, where $\sigma_0$ 
is the error estimated by the photometry pipeline, $\sigma_{\rm min}$ is a factor used to make the 
error to be consistent with the scatter of data, and $k$ is a scaling factor used to make $\chi^2$ 
per degree of freedom unity.  These factors are $(k, \sigma_{\rm min}/{\rm mag})=(1.432, 0.020)$, 
$(1.183, 0.015)$, and $(0.872,  0.020)$ for the KMTA, KMTC, and KMTS data sets, respectively.

\section{Interpretation of the anomaly}\label{sec:three}

\begin{figure}[t]
\includegraphics[width=\columnwidth]{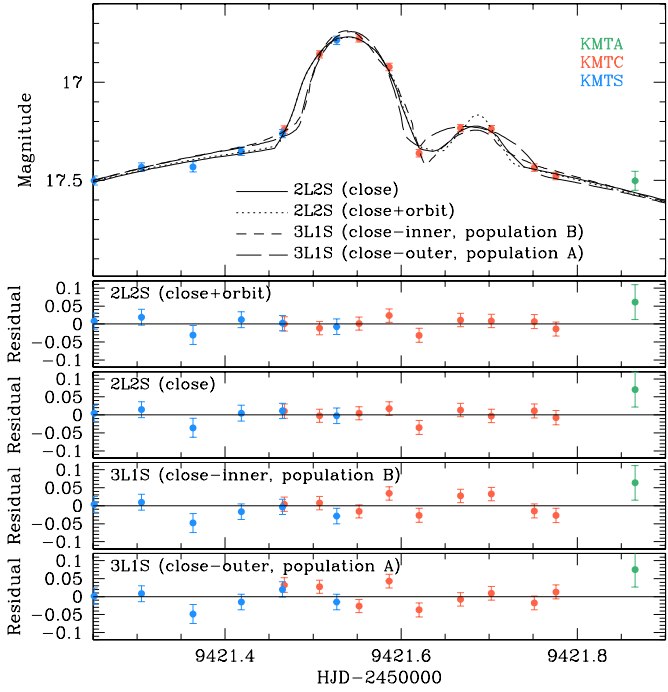}
\caption{
Zoom-in view around the peak of the light curve.  The four curves drawn over the data points are 
the two 2L2S and two 3L1S models.  One of the 2L2L (labeled as ``close+orbit'') model is obtained 
considering the orbital motion of the binary source, while the other model (``close'') is found 
without considering the source orbital motion.  The four lower panels show the residuals from the 
individual models.
}
\label{fig:two}
\end{figure}

Figure~\ref{fig:one} shows the light curve of KMT-2021-BLG-1898 constructed by combining the data 
sets from the three KMTNet telescopes.  Drawn over the data points is a model curve obtained from a 
single-lens single-source (1L1S) fit to the data.  As shown in the inset, the peak region exhibits 
a brief deviation from the 1L1S light curve.  The 1L1S fitting yields lensing parameters of 
$(t_0, u_0, \te)\sim (9421.5, 0.016, 22~{\rm days})$, where $t_0$ represents the time of the closest 
lens-source approach (expressed in ${\rm HJD}^\prime$), $u_0$ is the lens-source separation (scaled 
to the angular Einstein radius $\thetae$) at that time, and $\te$ represents the event time scale. 
An enlarged view of the peak region is presented in Figure~\ref{fig:two} to better show the detailed 
features of the anomaly. The anomaly, which lasted for about 8 hours, appears to be composed of two 
bumps centered at ${\rm HJD}^\prime \sim 9421.54$ (major bump) and $\sim 9421.68$ (minor bump). 
The major bump was covered by the combination of the KMTS and KMTC data sets, and the minor bump 
was covered by the KMTC data set.

% Figure 2 ------------------------------------------------------
% --------------------------------------------------------------

\subsection{2L1S and 1L2S interpretations}\label{sec:three-one}

% Table 1 ------------------------------------------------
\begin{table*}[t]
\small
%\tiny
%\centering
\caption{3L1S lensing parameters (population A)\label{table:one}}
\begin{tabular}{llllll}
%\begin{tabular*}{\columnwidth}{@{\extracolsep{\fill}}lcccc}
\hline\hline
\multicolumn{1}{c}{Parameter}       &
\multicolumn{1}{c}{close-outer}     &
\multicolumn{1}{c}{close-inner}     &
\multicolumn{1}{c}{wide-outer}      &
\multicolumn{1}{c}{wide-inner}      \\
\hline
$\chi^2$                &  $639.5             $  &  $640.2             $   &  $640.0             $   &  $640.0             $  \\
$t_0$ (HJD$^\prime$)    &  $9421.535 \pm 0.003$  &  $9421.535 \pm 0.003$   &  $9421.536 \pm 0.003$   &  $9421.535 \pm 0.003$  \\ 
$u_0$                   &  $0.015 \pm 0.001   $  &  $0.014 \pm 0.001   $   &  $0.014 \pm 0.001   $   &  $0.015 \pm 0.002   $  \\
$\te$ (days)            &  $23.87 \pm 2.33    $  &  $24.38 \pm 2.42    $   &  $24.06 \pm 2.23    $   &  $23.47 \pm 3.17    $  \\
$s_2$                   &  $0.776 \pm 0.008   $  &  $0.773 \pm 0.010   $   &  $1.288 \pm 0.012   $   &  $1.282 \pm 0.012   $  \\
$q_2$ ($10^{-3}$)       &  $1.192 \pm 0.172   $  &  $1.211 \pm 0.203   $   &  $1.097 \pm 0.143   $   &  $1.093 \pm 0.182   $  \\ 
$\alpha$ (rad)          &  $4.695 \pm 0.010   $  &  $4.696 \pm 0.010   $   &  $4.696 \pm 0.010   $   &  $4.692 \pm 0.011   $  \\
$s_3$                   &  $0.952 \pm 0.013   $  &  $1.068 \pm 0.015   $   &  $0.957 \pm 0.015   $   &  $1.053 \pm 0.015   $  \\
$q_3$ ($10^{-5}$)       &  $7.87 \pm 2.08     $  &  $8.10 \pm 1.78     $   &  $6.96 \pm 2.03     $   &  $6.56 \pm 2.15     $  \\  
$\psi$ (rad)            &  $5.792 \pm 0.010   $  &  $5.782 \pm 0.010   $   &  $5.795 \pm 0.011   $   &  $5.796 \pm 0.011   $  \\  
$\rho$ ($10^{-3}$)      &  $2.45 \pm 0.29     $  &  $2.50 \pm 0.28     $   &  $2.47 \pm 0.30     $   &  $2.56 \pm 0.38     $  \\
$I_{\rm S,KMT}$ (mag)   &  $21.83 \pm 0.11    $  &  $21.85 \pm 0.11    $   &  $21.84 \pm 0.11    $   &  $21.82 \pm 0.14    $  \\  
\hline
\end{tabular}
\tablefoot{ ${\rm HJD}^\prime = {\rm HJD}- 2450000$.  }
\end{table*}
% --------------------------------------------------------

% Table 2 ------------------------------------------------
\begin{table*}[t]
\small
%\tiny
%\centering
\caption{3L1S lensing parameters (population B)\label{table:two}}
\begin{tabular}{llllll}
%\begin{tabular*}{\columnwidth}{@{\extracolsep{\fill}}lcccc}
\hline\hline
\multicolumn{1}{c}{Parameter}       &
\multicolumn{1}{c}{close-outer}     &
\multicolumn{1}{c}{close-inner}      &
\multicolumn{1}{c}{wide-outer}      &
\multicolumn{1}{c}{wide-inner}      \\
\hline
$\chi^2$                &  $636.1                  $   &  $636.0                  $   &  $636.7                  $   & $637.7                   $   \\
$t_0$ (HJD$^\prime$)    &  $9421.536 \pm 0.003     $   &  $9421.536 \pm 0.003     $   &  $9421.534 \pm 0.003     $   & $9421.536 \pm 0.003      $   \\ 
$u_0$                   &  $0.015 \pm 0.002        $   &  $0.015 \pm 0.001        $   &  $0.015 \pm 0.001        $   & $0.016 \pm 0.002         $   \\
$\te$ (days)            &  $23.55 \pm 3.15         $   &  $23.42 \pm 1.87         $   &  $23.29 \pm 2.20         $   & $22.27 \pm 2.57          $   \\
$s_2$                   &  $0.750 \pm 0.026        $   &  $0.769 \pm 0.026        $   &  $1.313 \pm 0.032        $   & $1.309 \pm 0.049         $   \\
$q_2$ ($10^{-3}$)       &  $1.501 \pm 0.316        $   &  $1.464 \pm 0.269        $   &  $1.409 \pm 0.210        $   & $1.542 \pm 0.307         $   \\ 
$\alpha$ (rad)          &  $4.651 \pm 0.011        $   &  $4.645 \pm 0.010        $   &  $4.642 \pm 0.010        $   & $4.645 \pm 0.010         $   \\
$s_3$                   &  $0.986^{+0.001}_{-0.010}$   &  $0.997^{+0.010}_{-0.001}$   &  $0.985^{+0.002}_{-0.007}$   & $0.996^{+0.011}_{-0.001} $   \\
$q_3$ ($10^{-5}$)       &  $4.78 \pm 1.35          $   &  $5.45 \pm 1.25          $   &  $5.21 \pm 1.04          $   & $6.10 \pm 1.35           $   \\  
$\psi$ (rad)            &  $3.031 \pm 0.011        $   &  $3.042 \pm 0.011        $   &  $3.039 \pm 0.011        $   & $3.051 \pm 0.012         $   \\  
$\rho$ ($10^{-3}$)      &  $2.78 \pm 0.34          $   &  $2.82 \pm 0.24          $   &  $2.82 \pm 0.25          $   & $2.87 \pm 0.31           $   \\
$I_{\rm S,KMT}$ (mag)   &  $21.82 \pm 0.14         $   &  $21.81 \pm 0.09         $   &  $21.81 \pm 0.10         $   & $21.75 \pm 0.11          $   \\  
\hline
\end{tabular}
%\tablefoot{ ${\rm HJD}^\prime = {\rm HJD}- 2450000$.  }
\end{table*}
% --------------------------------------------------------

To investigate the origin of the anomaly, we first conducted modeling of the observed light curve 
under the two interpretations that the lens is a binary (2L1S model) in one interpretation and 
the source is a binary (1L2S model) in the other interpretation.
Both the 2L1S and 1L2S models require one to include extra parameters in addition to those 
of the 1L1S model, that is, $(t_0, u_0, \te)$.  These extra parameters for the 2L1S model
are $(s, q, \alpha, \rho)$,  which represent the projected separation (scaled to $\thetae$) 
and mass ratio between the binary lens component, $M_1$ and $M_2$, the angle between the 
source motion and the binary lens axis (source trajectory angle), and the ratio of the angular 
source radius $\theta_*$ to $\thetae$ (normalized source radius), respectively.  The normalized 
source radius is needed to account for the deformation of an anomaly caused by finite-source 
effects because 2L1S light curves are often involved with caustics \citep{Bennett1996}.  
The extra parameters for the 1L2S model include $(t_{0,2}, u_{0,2}, \rho_2, q_F)$, which 
indicate the closest approach time, separation, and normalized radius of the second source, and 
the flux ratio between the binary source stars, $S_1$ and $S_2$, respectively \citep{Hwang2013}.

The modeling was done using the combination of a grid search and a downhill approach. In
the 2L1S modeling, we divided the lensing parameters into two groups, in which the parameters
$s$ and $q$ of the first group were searched for using a grid approach, while the other
parameters in the second group were found using a downhill approach. We applied the Markov
Chain Monte Carlo (MCMC) method for the downhill approach.  We identified local solutions from 
the inspection of the $\Delta\chi^2$ map on the $\log s$--$\log q$ parameter plane constructed 
from the grid search, and we then refined the individual local solutions by releasing all parameters 
as free parameters. For the 1L2S model, all parameters were searched for using a downhill approach, 
in which the initial values of the lensing parameters were given considering the times and magnitudes 
of the anomaly features.  From the modeling under the 2L1S and 1L2S interpretations, it was found 
that neither of the models could explain the double-bump features of the anomaly, although both 
models could approximately describe one of the two bumps.

% Figure 3 ------------------------------------------------------
\begin{figure}[t]
\includegraphics[width=\columnwidth]{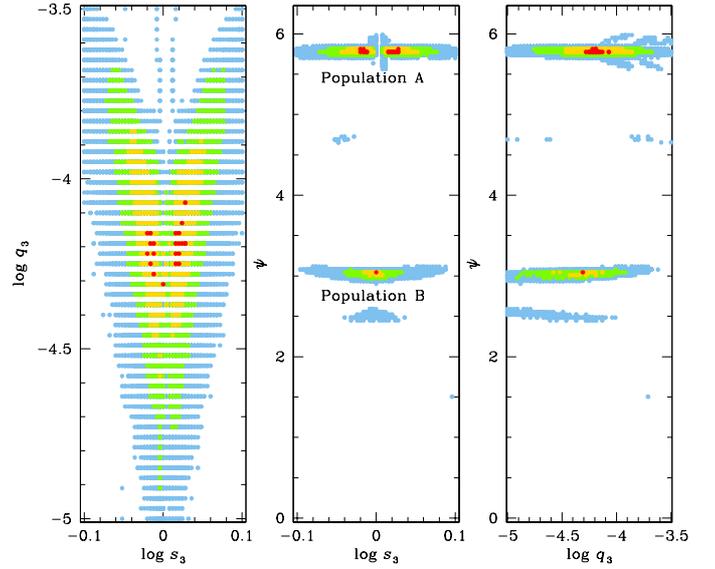}
\caption{
Maps of $\Delta\chi^2$ on the $\log s_3$--$\log q_3$--$\psi$ parameter planes obtained from the 
3L1S modeling. Color coding is set to represent points with $\Delta\chi^2\leq 1n\sigma$ (red), 
$\leq 2n\sigma$  (yellow), $\leq 3n\sigma$  (green), and $\leq 4n\sigma$ (cyan), where $n=2$.
}
\label{fig:three}
\end{figure}
% --------------------------------------------------------------

\subsection{3L1S interpretation}\label{sec:three-two}

Recognizing the inadequacy of the 2L1S and 1L2S models in describing the anomaly, we first 
checked the interpretation that the lens is composed of three masses ($M_1$, $M_2$, and $M_3$).  
We examined the 3L1S interpretation due to the combination of the two facts; first, one of the 
bumps can be explained by a 2L1S model with a planet-mass lens companion, and second, the 
anomaly appears near the peak of the light curve.  If the lens contains a tertiary component, 
which can be either a second planet lying in the vicinity of the Einstein ring \citep{Gaudi1998} 
or a binary companion to the host of the planet with a very close or a wide separation from the 
host \citep{Lee2008}, the tertiary lens component induce an extra caustic in the central 
magnification region, and the second bump, which could not be described with a single-planetary 
model, may be explained by a 3L1S model.

For a 3L1S modeling, extra parameters are needed in addition to those of a 2L1S modeling. 
These extra parameters are $(s_3, q_3, \psi)$, which represent the separation and mass ratio 
between $M_1$ and $M_3$, and the orientation angle of $M_3$ as measured from the $M_1$--$M_2$ 
axis with a center at the position of $M_1$. In the 3L1S model, we designate the parameters 
related to $M_2$ as $(s_2, q_2)$ to distinguish them from the parameters related to $M_3$.

It is known that anomalies in the central magnification region can often be approximately described 
by the superposition of the anomalies induced by the two $M_1$--$M_2$ and $M_1$--$M_3$ binary pairs 
\citep{Bozza1999, Han2001}. Under this approximation, we conducted a 3L1S modeling according to 
the following procedure. In the first step, we conducted a grid searches for $(s_3, q_3, \psi)$ 
by fixing the other lensing parameters as the values of the 2L1S model describing the major bump. 
We checked local solutions on the three $\log s_3$--$\log q_3$--$\psi$ parameter planes, and then 
polished the individual local solutions identified from the first-round grid search by allowing 
all parameters to vary.

\begin{figure}[t]
\includegraphics[width=\columnwidth]{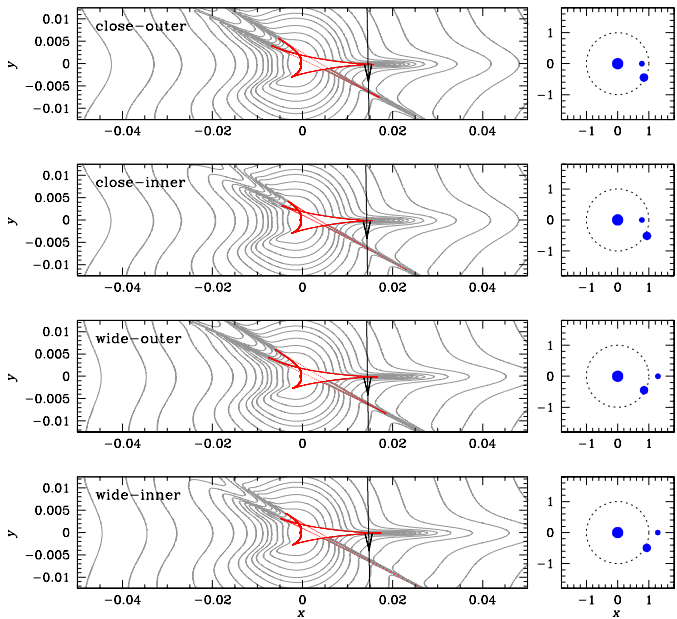}
\caption{
Lens system configurations of the four 3L1S models in the population A. For each model, 
the left panel shows the caustic (red cuspy figure) and source trajectory (line with an 
arrow) in the central magnification region, and the right panel shows the positions of the 
lens components (marked by blue filled dots) around the Einstein ring (dotted circle).  
The sizes of the blue dots are set according to the order of masses of the lens components. 
Lengths are scaled to the Einstein radius corresponding to the total mass of the lens.
The grey curves encompassing the caustic represent the equi-magnification contours. 
}
\label{fig:four}
\end{figure}

\begin{figure}[t]
\includegraphics[width=\columnwidth]{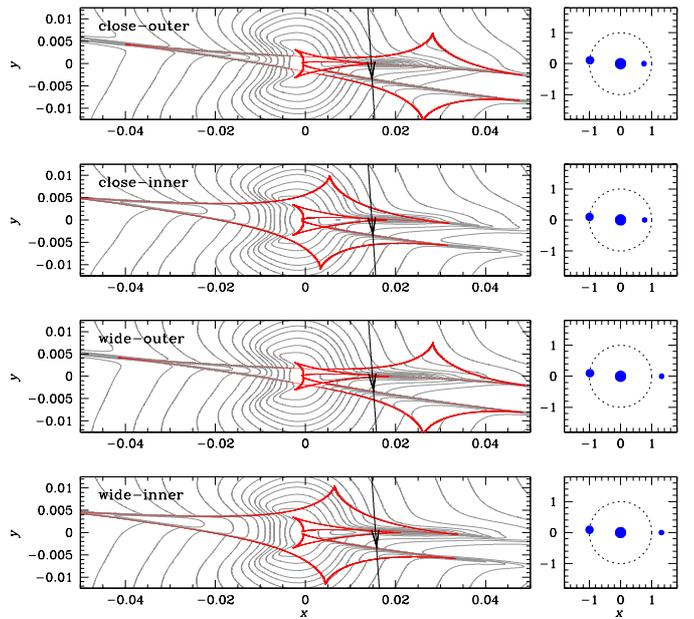}
\caption{
Lens system configurations of the 3L1S models in the population B. Notations are same as
those in Fig.~\ref{fig:four}.
}
\label{fig:five}
\end{figure}

Figure~\ref{fig:three} shows the $\Delta\chi^2$ maps on the $\log s_3$--$\log q_3$--$\psi$ 
parameter planes obtained from the grid search.  We identified two distinctive populations 
of local solutions caused by the accidental degeneracy in determining $\psi$: solutions in 
``population A'' with $\psi\sim 330^\circ$ and solutions in ``population B'' with 
$\psi\sim 173^\circ$.  For each population, there are four degenerate solutions resulting from 
degeneracies in determining the separations $s_2$ ($M_1$--$M_2$ degeneracy) and $s_3$ ($M_1$--$M_3$ 
degeneracy), and thus there exist eight solutions in total.  These degeneracies will be further 
discussed below.  Figures~\ref{fig:four} and \ref{fig:five} show the lens system configurations 
of the individual 3L1S models in the A and B populations, respectively.

% Figure 4 ------------------------------------------------------
% --------------------------------------------------------------

The degeneracy between the solutions in the populations A and B is ``accidental'' in the following 
sense. From Figures~\ref{fig:four} and \ref{fig:five}, we can see that the $M_1$-$M_3$ caustic would, 
by itself, give rise to a single bump shortly after $t_0$ for population A, but would give rise to 
two bumps (one at $t_0$ and the other shortly after) for population B. However, the first bump in 
population B, which (like the second bump) is weak, is superposed on the main bump that is generated 
by the $M_1$--$M_2$ caustic. Hence, its impact cannot be distinguished given the cadence and quality 
of the data, and it is rather absorbed into the $M_1$--$M_2$ model parameters. Similarly, 
Figure~\ref{fig:two} shows that higher cadence over the second bump would have easily distinguished 
between populations A and B.

For convenience and clarity, we label the $M_1$--$M_2$ degeneracy as ``close-wide'' and the $M_1$--$M_3$
degeneracy as ``outer-inner''.  The outer-inner degeneracy was originally proposed by \citet{Gaudi1997} 
for trajectories going ``outside'' and ``inside'' a planetary caustic.  \citet{Hwang2022} pointed out 
that in the limit of trajectories passing near the planetary caustic, the separations $s_\pm$ (for 
inner and outer) obey the relation
\begin{equation}
{ 1\over 2} (s_+ + s_-) = s^\dagger, 
\label{eq1}
\end{equation}
where
\begin{equation}
s^\dagger = {\sqrt{u_{\rm anom}^2 + 4} \pm u_{\rm anom}\over 2},
\qquad
u_{\rm anom}^2 = u_0^2 + {(t_{\rm anom} - t_0)^2\over t_{\rm E}^2},
\label{eq2}
\end{equation}
and $t_{\rm anom}$ represents the time of the anomaly, with the sign ``$\pm$'' applying to ``bump'' 
(positive) and ``dip'' (negative) anomalies, respectively.  Gould et al. (2022, in preparation) 
argued that this could be generalized for trajectories that were not in the immediate neighborhood 
of the planetary caustic to
\begin{equation}
\sqrt{s_+ s_-} = s^\dagger \qquad ({\rm outer-inner}).
\label{eq3}
\end{equation}

\citet{Griest1998} derived a close-wide degeneracy in the limit of planetary anomalies
near the peak of high-magnification events, that is, $u_{\rm anom}\rightarrow 0$, for which they
showed that $s_- = 1/s_+$. We write this didactically (i.e., with an ``unnecessary'' square-root
symbol) as
\begin{equation}
\sqrt{s_+ s_-} = 1 \qquad ({\rm close-wide}).
\label{eq4}
\end{equation}
With rare exceptions, virtually all degeneracy pairs were referred to in the literature as 
``close-wide'', even when they did not obey this relation, even approximately.  \citet{Herrera2020} 
first noticed that one such ``nonobeying'' case was in fact the outer-inner degeneracy, with the 
outer solution having smaller $s$, and so being incorrectly labeled ``close''.  \citet{Yee2021} 
then argued that the transition between the two types of degeneracies is continuous. That is, 
it passes continuously from the close-wide limit of central caustics, through resonant caustics, 
to the outer-inner limit of planetary caustics. In retrospect, we can see that Equation~(\ref{eq4}) 
is a special case of Equation~(\ref{eq3}) because in the close-wide limit, $u_{\rm anom}\rightarrow 0$, 
so $s^\dagger \rightarrow 1$. That is, Equation~(\ref{eq3}) gives a mathematical expression to the 
unification conjecture of \citet{Yee2021}.

% Figure 5 ------------------------------------------------------
% --------------------------------------------------------------

In the present case, population B should clearly be labeled ``outer-inner'' because the two values
of $s_3$ are both less than unity.  For population A, the $M_1$--$M_3$ geometry is equally far from 
the limits in which the outer-inner and close-wide degeneracies were derived, and so could be 
referred to as either. We choose to call them ``outer-inner'' in order to maintain the most consistent 
notation.

It is found that the 3L1S models can approximately describe the double-bump features of the 
anomaly.  In Figure~\ref{fig:two}, we present the model curves and residuals of the best-fit 
solutions in the A (``close-outer'' model) and B (``close-inner'' model) populations.  In 
Tables~\ref{table:one} and \ref{table:two}, we list the lensing parameters of the 3L1S solutions 
in the populations ``A'' and ``B,'' respectively, along with $\Delta\chi^2$ values of the model 
fits and the magnitude of the source according to the KMTNet scale, $I_{\rm S,KMT}$.  The 
close-inner model of the population~B solutions provides the best fit to the data, but the 
$\Delta\chi^2$ differences relative to the other models are $\Delta\chi^2\leq 4.2$, indicating 
that the degeneracies among the solutions are severe.

\subsection{2L2S interpretation}\label{sec:three-three}

The double bump feature of the anomaly may be depicted if the source is a binary and if the second 
source additionally approached or crossed the caustic induced by the $M_1$--$M_2$ binary lens 
system.  We checked this possibility by conducting a model in which both the lens and source are 
binaries (2L2S model). In addition to the parameters of a 2L1S model, a 2L2S model requires one 
to include four extra parameters of $(t_{0,2}, u_{0,2}, \rho_2, q_F)$, where the subscript ``2'' 
designate the second source. We use the subscript ``1'' to designate the corresponding parameters 
related to the primary source, that is, $(t_{0,1}, u_{0,1}, \rho_1)$. In the 2L2S modeling, we 
started with the lensing parameters of the 2L1S model describing the major bump, and tested various 
trajectories of the second source to check whether the minor bump could be explained by the second 
source.

\begin{figure}[t]
\includegraphics[width=\columnwidth]{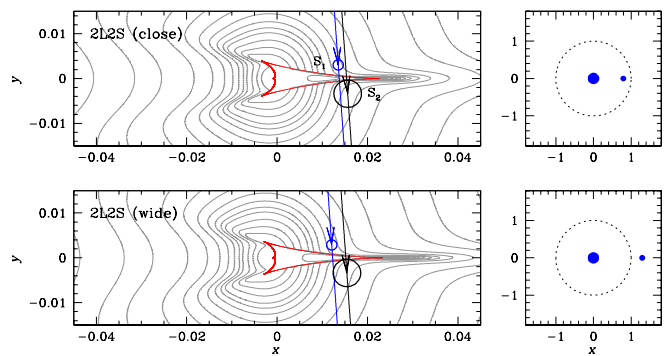}
\caption{
Lens system configurations of the close (upper panel) and wide (lower panel) 2L2S solutions. 
Notations are same as those in Fig.~\ref{fig:four}, except that there are two lens masses and 
two source trajectories. The source trajectories of the first and second source stars  are 
marked by $S_1$ and $S_2$, respectively. The empty circles on the source trajectories represent 
the source positions at a certain epoch, and the sizes of the circles are scaled to the caustic 
size.  
}
\label{fig:six}
\end{figure}

From the 2L2S modeling, we found two solutions that could depict the double-bump feature of the
anomaly. The two solutions were found from the two sets of the initial lensing parameters adopted
from the close and wide 2L1S solutions, and we designate the individual solutions with $s<1.0$ and
$s>1.0$ as the ``close'' and ``wide'' solutions, respectively. The lensing parameters and $\chi^2$ 
values of the two 2L2S solutions are listed in Table~\ref{table:three}, and the corresponding lens 
system configurations are shown in Figure~\ref{fig:six}. It is found that the close model yields a 
better fit than the wide model, but the $\chi^2$ difference between the models, $\Delta\chi^2=0.2$, 
is very small. The model curve and the residual of the close solution are shown in Figure~\ref{fig:two}.  
According to the lensing configurations, both the major and minor bumps of the anomaly were produced 
by the successive crossings of the binary source stars over the central caustic induced by a planetary 
companion to the lens. The flux from the second source, which trailed the first source and approached 
the caustic more closely than the primary source, comprises $\sim 17\%$ of the $I$-band flux from the 
first source. From the comparison of the $\chi^2$ values with those of the 3L1S solutions, it is found 
that the 2L2S solutions yield a better fit than the 3L1S solutions with $\Delta\chi^2=[11.1$--15.3].

% Table 3 ------------------------------------------------
\begin{table}[t]
\small
%\centering
\caption{2L2S lensing parameters\label{table:three}}
\begin{tabular*}{\columnwidth}{@{\extracolsep{\fill}}llll}
\hline\hline
\multicolumn{1}{c}{Parameter}    &
\multicolumn{1}{c}{Close}        &
\multicolumn{1}{c}{Wide }        \\
\hline
$\chi^2$                  &  $624.9             $  & $625.1             $   \\
$t_{0,1}$ (HJD$\prime$)   &  $9421.518 \pm 0.005$  & $9421.517 \pm 0.004$   \\ 
$u_{0,1}$                 &  $0.016 \pm 0.002   $  & $0.015 \pm 0.002   $   \\
$t_{0,2}$ (HJD$\prime$)   &  $9421.669 \pm 0.009$  & $9421.667 \pm 0.008$   \\
$u_{0,2}$                 &  $0.012 \pm 0.004   $  & $0.012 \pm 0.003   $   \\
$\te$ (days)              &  $22.61 \pm 2.21    $  & $23.264 \pm 2.21   $   \\ 
$s$                       &  $0.795 \pm 0.018   $  & $1.297 \pm 0.029   $   \\
$q$ ($10^{-3}$)           &  $1.46 \pm 0.29     $  & $1.57 \pm 0.25     $   \\
$\alpha$ (rad)            &  $4.648 \pm 0.014   $  & $4.643 \pm 0.013   $   \\
$\rho_1$ ($10^{-3}$)      &  $3.12 \pm 0.33     $  & $2.98 \pm 0.30     $   \\ 
$\rho_2$ ($10^{-3}$)      &  $1.70 \pm 0.64     $  & $1.68 \pm 0.61     $   \\
$q_F$                     &  $0.17 \pm 0.05     $  & $0.17 \pm 0.05     $   \\
$I_{\rm S,KMT}$ (mag)     &  $21.77 \pm 0.12    $  & $21.80 \pm 0.10    $   \\
\hline
\end{tabular*}
%\tablefoot{ ${\rm HJD}^\prime = {\rm HJD}- 2450000$.  }
\end{table}
% --------------------------------------------------------

According to the static 2L2S solutions, the separation between the two source stars at the time of 
the anomaly, $\Delta u =\{ [(t_{0,1}-t_{0,2})/\te]^2+(u_{0,1}-u_{0,2})^2\}^{1/2} \sim 6.7\times 
10^{-3}$, is very small. Assuming that this corresponds to the semi-major axis of the source orbit, 
that is, $a\sim \Delta u D_{\rm S}\thetae\sim 0.02$~AU, and the masses of the binary source stars 
are $M_{S_1}=1~M_\odot$ and $M_{S_2}=0.5~M_\odot$, the orbital period of the binary source is 
$P=[(a/{\rm AU})^3/(M_S/M_\odot)]^{1/2}\sim 1$~day, where $M_S=M_{S_1}+M_{S_2}$.  Because this 
orbital period is of the same order as the duration of the anomaly, the orbital motion of the binary 
source may be important to the binary-source modeling. Although it would be difficult to define the 
orbital lensing parameters based on the handful of data points covering the short-term anomaly, we 
conducted an additional modeling considering the source orbital motion to check whether the fit 
further improves with the consideration of the source orbital motion. From the modeling conducted 
under the assumption of a simplified face-on circular orbit, it is found that the fit improves by 
$\Delta\chi^2\sim 3.2$ with respect to the static model, making the gap between the 2L2S and 3L1S 
solutions wider, into $\Delta\chi^2 =[14.3$--18.5]. The orbital 2L2S model and its residual are 
shown in Figure~\ref{fig:two}.

% Figure 6 ------------------------------------------------------
% --------------------------------------------------------------

As is discussed in the following section, the Einstein radii derived independently from the 
two stars of the binary source result in consistent values.  Together with the better fit, we 
conclude that the single-planet binary-source 2L2S interpretation of the anomaly is more 
plausible than the multi-planet 3L1S interpretation.  According to the 2L2S interpretation, 
KMT-2021-BLG-1898 is the fifth binary-lensing event occurring on a binary stellar system, 
following MOA-2010-BLG-117 \citep{Bennett2018}, OGLE-2016-BLG-1003 \citep{Jung2017}, 
KMT-2018-BLG-1743 \citep{Han2021a}, and KMT-2019-BLG-0797 \citep{Han2021b}. For three of 
these events (MOA-2010-BLG-117, KMT-2018-BLG-1743, and KMT-2021-BLG-1898), the lenses are 
planetary systems.

\section{Source stars and Einstein radius}\label{sec:four}

We specify the source not only to fully characterize the event but also to measure the angular
Einstein radius. The source was specified by measuring its color and magnitude. According to the
routine procedure, the first step for this specification is measuring the source magnitudes in 
two passbands, $I$ and $V$ bands in our case, from the regression of the photometric data to the 
lensing model.  For  KMT-2021-BLG-1898, the $I$-band source magnitude was precisely measured, but 
a reliable measurement of the $V$-band magnitude was difficult because the quality of the $V$-band 
data was not good due to the heavy extinction toward the field.  We, therefore, estimated the 
source color by interpolating it from the main-sequence (MS) branch of stars in the CMD constructed 
from the {\it Hubble Space Telescope} ({\it HST}) observations \citep{Holtzman1998}.

The detailed procedure of specifying the source type is as follows. First, we estimated the combined 
$I$-band source flux, $F_{S,I}=F_{S_1,I}+F_{S_2,I}$, from the regression of the $I$-band data processed 
using the pyDIA code to the model.  Here $F_{S_1,I}$ and $F_{S_2,I}$ represent the $I$-band flux values 
from the primary and secondary source stars, respectively.  With the flux ratio $q_{F,I}$ between the 
two source stars estimated from the modeling, we then estimated the flux values of the individual 
source stars as
\begin{equation}
F_{S_1,I} = \left( {1 \over 1+q_{F,I}} \right)F_{S,I};\qquad 
F_{S_2,I} = \left( {q_{F,I} \over 1+q_{F,I}} \right)F_{S,I}. 
\label{eq5}
\end{equation}
Second, we made a combined CMD by aligning the {\it HST} CMD and that constructed with the pyDIA 
photometry of the KMTC data set using the centroids of red giant clumps (RGCs) in the individual 
CMDs.  We then estimate the colors of $S_1$ and $S_2$ by interpolating them from the MS branch on 
the {\it HST} CMD corresponding to the $I$-band magnitudes of $S_1$ and $S_2$.

\begin{figure}[t]
\includegraphics[width=\columnwidth]{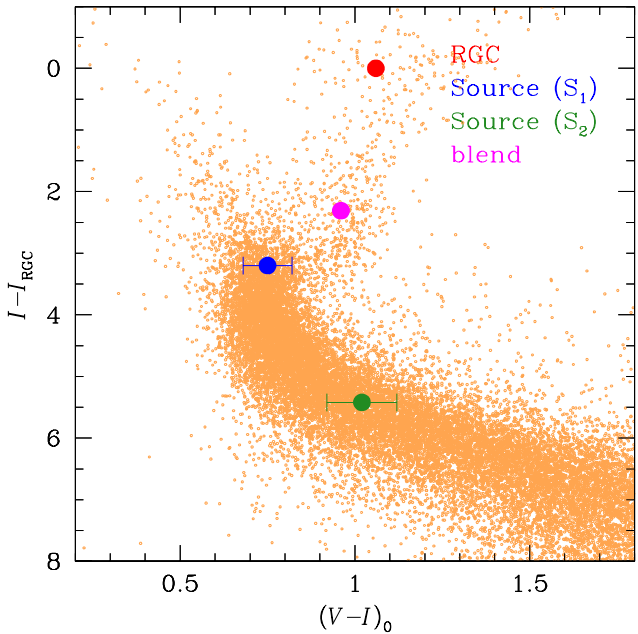}
\caption{
Locations of the primary ($S_1$) and secondary ($S_2$) source stars with respect to the centroid 
of red giant clump (RGC) in the color-magnitude diagram constructed from the {\it Hubble Space 
Telescope} observations.  The abscissa represents the de-reddened color $(V-I)_0$, while the 
ordinate denotes the $I$-band magnitude difference with respect to that of the RGC centroid.
}
\label{fig:seven}
\end{figure}

It is found that the two source stars are 3.20~mag and 5.43~mag fainter than the RGC centroid.  
With the known values of the extinction and reddening-corrected (de-reddened) color and magnitude, 
$(V-I, I)_{\rm RGC}=(1.06, 14.39)$ \citep{Bensby2013, Nataf2013}, for the RGC centroid, we estimate 
that the de-reddened colors and magnitudes of the source stars are 
\begin{equation}
(V-I, I)_{0,S_1}=(0.75\pm 0.07, 17.59\pm 0.05)
\label{eq6}
\end{equation}
for the primary source, and 
\begin{equation}
(V-I, I)_{0,S_2}=(1.02\pm 0.10, 19.81\pm 0.40) 
\label{eq7}
\end{equation}
for the secondary source.  Here, the error of the $I$-band magnitude for $S_2$ was estimated from 
the error propagation of the magnitude uncertainty of $S_1$ together with the uncertainty of 
$q_{F,I}$ measurement.  For each star, we estimated the color directly from the median color of 
stars with the same offset from the clump on the {\it HST} CMD based on images of Baade's window.  
We derived the error bars from the scatter in $(V-I)$ at fixed offset by first taking account of 
the photometric measurement errors that were described by \citet{Holtzman1998}.  Figure~\ref{fig:seven} 
shows the locations of $S_1$ and $S_2$ on the {\it HST} CMD.  The estimated colors and magnitudes 
indicate that the primary source is a turnoff star or a subgiant with a G spectral type, and the 
secondary source is a mid K-type dwarf.  Also marked in the CMD is the location of the blend, which 
is fainter than the RGC centroid by 1.25~mag in the $I$ band.  As was in the case of the source, 
the color of the blend could not be measured directly from the photometric data due to the poor 
$V$-band data.  We, therefore, estimate the blend color as the median value of the giant branch 
corresponding to the $I$-band magnitude difference from the RGC centroid.

% Figure 7 ------------------------------------------------------
% --------------------------------------------------------------

With the specified source type, we estimate the angular Einstein radius by
\begin{equation}
\thetae = {\theta_{*,S_1}\over \rho_1}
\label{eq8}
\end{equation}
where $\theta_{*,S_1}$ represents the angular radius of $S_1$.  For the estimation of the angular 
source size from the measured color, we first convert $V-I$ color into $V-K$ using the color-color 
relation of \citet{Bessell1988}, and then interpolate $\theta_{*,S_1}$ from the $(V-K)$--$\theta_*$ 
relation of \citet{Kervella2004}. This yields the angular radius of the primary source of
\begin{equation}
\theta_{*,S_1} = 0.99 \pm 0.11~\mu{\rm as}, 
\label{eq9}
\end{equation}
and the angular Einstein radius of
\begin{equation}
\thetae = 0.32 \pm 0.05~{\rm mas}. 
\label{eq10}
\end{equation} The relative lens-source proper motion is estimated from the combination of the Einstein
radius and event time scale as
\begin{equation}
\mu = {\thetae\over \te}  = 5.11 \pm 0.47~{\rm mas}~{\rm yr}^{-1}.
\label{eq11}
\end{equation}

\begin{figure}[t]
\includegraphics[width=\columnwidth]{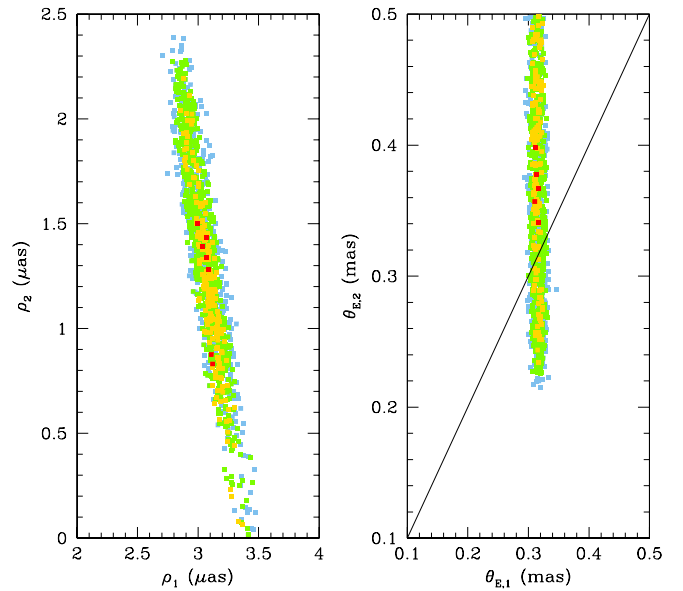}
\caption{
Scatter plot of points in the MCMC chain on the $\rho_1$--$\rho_2$ (left panel) and 
$\theta_{\rm E,1}$--$\theta_{\rm E,2}$ (right panel) planes obtained from the 2L2S modeling.  
The color coding is the same as that in Fig.~\ref{fig:two} except that $n=1$. The diagonal line 
in the right panel represents the relation $\theta_{\rm E,1} = \theta_{\rm E,2}$.
}
\label{fig:eight}
\end{figure}

The validity of the 2L2S interpretation is further supported by the fact that the angular Einstein 
radii derived independently from the two stars of the binary source result in consistent values.  
To demonstrate this, we estimated two values of $\thetae$: one derived from the source type and 
$\rho_1$ of the primary source, $\theta_{\rm E,1}$, and the other from those of the secondary source,  
$\theta_{\rm E,2}$.  Figure~\ref{fig:eight} shows the scatter plots of MCMC points on the 
$\rho_1$--$\rho_2$ (left panel) and $\theta_{\rm E,1}$--$\theta_{\rm E,2}$ (right panel) planes.  
We note that the $\theta_{\rm E,1}$--$\theta_{\rm E,2}$ scatter plot is elongated along the ordinate 
direction due to the large uncertainty of $\rho_2$.  The plots shows that the Einstein radii estimated 
from $S_1$ and $S_2$ result in consistent values, which is $\thetae\sim 0.3$~mas, and this further 
supports the 2L2S model in addition to its better fit than fit of the 3L1S interpretation.

% Figure 8 ------------------------------------------------------
% --------------------------------------------------------------

% Table 4 ------------------------------------------------
\begin{table}[t]
\small
%\centering
\caption{Physical lens parameters\label{table:four}}
\begin{tabular*}{\columnwidth}{@{\extracolsep{\fill}}lll}
\hline\hline
\multicolumn{1}{c}{Quantity}      &
\multicolumn{1}{l}{Close}         &
\multicolumn{1}{l}{Wide}          \\
\hline
$M_{\rm host}$ ($M_\odot$)        &   $0.48^{+0.33}_{-0.26}$  &  $\leftarrow$               \\  [0.8ex]
$M_{\rm planet}$ ($M_{\rm J}$)    &   $0.73^{+0.50}_{-0.40}$  &  $0.79^{+0.54}_{-0.43}$     \\  [0.8ex]
$\dl$ (kpc)                       &   $6.9^{+0.9}_{-1.3}   $  &  $\leftarrow$               \\  [0.8ex]
$a_{\perp}$ (AU)                  &   $1.9^{+0.3}_{-0.3}   $  &  $3.0^{+0.4}_{-0.6}$        \\  [0.8ex]
\hline                           
\end{tabular*}
\tablefoot{
The arrows in the third column indicate that the values are same as those in the second column.}
\end{table}
% --------------------------------------------------------

\section{Physical parameters of the planetary system}\label{sec:five}

The lensing observables that can constrain the physical lens parameters of the mass, $M$, and
distance, $\dl$, include the event time scale $\te$, Einstein radius $\thetae$, and microlens 
parallax $\pie$. With the measurements of all these observables, the physical parameters can be 
uniquely determined as
\begin{equation}
M= {\thetae\over \kappa \pie};\qquad
\dl = {{\rm AU}\over \pie\thetae+ \pi_{\rm S}},
\label{eq12}
\end{equation}
where $\kappa=4G/(c^2{\rm AU})$, $\pi_{\rm S}={\rm AU}/D_{\rm S}$ is the trigonometric parallax of 
the source lying at a distance $D_{\rm S}$ \citep{Gould2000}. For KMT-2021-BLG-1898, the observables 
of $\te$ and $\thetae$ were measured from the modeling of the light curve. However, the microlens 
parallax $\pie$, which can be measured from the subtle deviations in the lensing light curve from 
the symmetric form induced by the positional change of the source caused by the orbital motion of 
Earth around the Sun \citep{Gould1992}, could not be securely measured due to the relatively short 
time scale, $\te\sim 23$~days, of the event together with the low precision of the photometric
 data. Although $M$ and $\dl$ cannot be uniquely determined, one can still constrain them using 
the other observables, which are related to the physical parameters by
\begin{equation}
t_{\rm E} = {\thetae \over \mu};\qquad
\thetae = (\kappa M \pi_{\rm rel})^{1/2},
\label{eq13}
\end{equation}
where $\pi_{\rm rel}= {\rm AU}(D_{\rm L}^{-1}-D_{\rm S}^{-1})$ is the relative lens-source parallax.

% Figure 9 ------------------------------------------------------
\begin{figure}[t]
\includegraphics[width=\columnwidth]{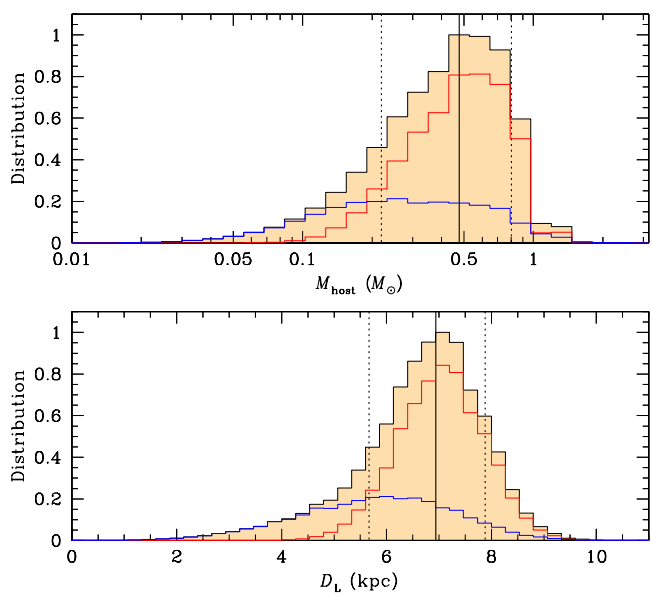}
\caption{
Bayesian posteriors of the host mass (upper panel) and distance (lower panel) to the planetary system. 
In each panel, the red and blue curves represent the contributions by the bulge and disk lens populations, 
respectively, and the black curve is the sum of the two populations. The solid vertical line indicates 
the median of the distribution, and the two dotted vertical lines represent the 1$\sigma$ range of the 
distribution.
}
\label{fig:nine}
\end{figure}
% --------------------------------------------------------------

The lens parameters were estimated from a Bayesian analysis conducted with the use of the measured
observables $\te$ and $\thetae$. In the first step of the Bayesian analysis, we conducted a Monte Carlo
simulation to produce a large number ($2\times 10^6$) of artificial lensing events. The simulation 
was done using a prior Galactic model defining the locations, motions, and masses of Galactic objects. 
We used the Galactic model of \citet{Jung2021} in the simulation. The model was constructed using
the \citet{Robin2003} and \citet{Han2003} models for the physical distributions of disk and bulge 
objects, respectively, \citet{Jung2021} and \citet{Han1995} models for the dynamical distributions 
of the disk and bulge objects, respectively, and \citet{Jung2018} model for the mass function for 
the lenses. In the second step, we constructed the posterior distributions of the physical lens 
parameters of $M$ and $\dl$ for the simulated events with event time scales and angular Einstein 
radii that were consistent with observed values of $\te$ and $\thetae$. With the constructed 
distributions, we present the median values as the representative values of the physical parameters, 
and set the 16\% and 84\% of the distributions as the lower and upper limits of the 1$\sigma$ ranges.

Figure~\ref{fig:nine} shows the posterior distributions of the mass of the planet host and distance 
to the planetary system constructed from the Bayesian analysis. In Table~\ref{table:four}, we list 
the estimated parameters of the host mass, $M_{\rm host}$, planet mass, $M_{\rm planet}$, distance, 
and projected separation between the planet and host, $a_\perp=s\dl \thetae$, corresponding to the 
close and wide solutions. It turns out that the lens is a planetary system in which a planet with 
a mass of $\sim 0.7$--0.8~$M_{\rm J}$ orbits an early M dwarf host with a mass of $\sim 0.5~M_\odot$, 
and the projected planet-host separation is $\sim 1.5$~AU and $\sim 2.5$~AU according to the close 
and wide solutions, respectively. The relative probabilities for the lens to be in the disk and 
bulge are 32\% and 68\%, respectively, and thus the planetary system is more likely to be in the 
bulge.  According to the estimated mass and distance, the $I$-band magnitude of the planet host is 
$I\sim M_I + 5 \log \dl -5 + A_I \sim 25.5$, where $M_I\sim 7.7$ is the $I$-band absolute magnitude 
corresponding to the mass.  Due to the faintness, the lens does not appear in the list of the Gaia 
Early Data Release 3 \citep{Gaia2021}.

\section{Conclusion}\label{sec:six}

We analyzed the microlensing event KMT-2021-BLG-1898, for which the lensing light curve exhibited 
a short-term anomaly near the peak.  It was found that the anomaly with double-bump features could 
not be explained by the usual binary-lens or binary-source interpretations.  In order to reveal 
the nature of the anomaly, we conducted modeling of the light curve under various sophisticated 
models  with the inclusion of additional lens or source component.

We found that the anomaly was best explained by a model, in which both the lens and source are 
binaries.  For this model, the lens is a planetary system with a planet/host mass ratio, and the 
source is a binary composed of a turn-off or a subgiant star and a mid K dwarf.  The double-bump 
feature of the anomaly could also be described by a triple-lens model, in which the lens is a 
planetary system containing two planets.  Among the two models, the 2L2S model was favored over 
the 3L1S model not only because it yields a better fit to the data but also the Einstein radii 
derived independently from the two stars of the binary source result in consistent values.  
According to the 2L2S interpretation, KMT-2021-BLG-1898 is the fifth event in which both the lens 
and source are binaries, and the third 2L2S case in which the lens is a planetary system.  degeneracy 
in the planet-host separation.

From a Bayesian analysis based on the observables of the event, we estimated that the planet has 
a mass of $\sim 0.7$--0.8~$M_{\rm J}$, and it orbits an early M dwarf host with a mass of 
$\sim 0.5~M_\odot$.  The projected planet-host separation is $\sim 1.5$~AU and $\sim 2.5$~AU 
according to the close and wide solutions, respectively.

\begin{acknowledgements}
Work by C.H. was supported by the grants  of National Research Foundation of Korea 
(2020R1A4A2002885 and 2019R1A2C2085965).
J.C.Y. acknowledges support from N.S.F Grant No. AST-2108414.
% KMTNet
This research has made use of the KMTNet system operated by the Korea Astronomy and Space 
Science Institute (KASI) and the data were obtained at three host sites of CTIO in Chile, 
SAAO in South Africa, and SSO in Australia.
\end{acknowledgements}

\end{document}